
\baselineskip16.5pt
\abovedisplayskip 6pt plus2pt minus5pt
\belowdisplayskip 6pt plus2pt minus5pt
\vsize=9.7 true in \raggedbottom
\hsize=6.5 in
\voffset= -0.2 true in
\font\bmit=cmmib10 \textfont9=\bmit \def\bmit{\fam9 }
\font\bx=cmbsy10 \textfont10=\bx \def\bx{\fam10 }

\mathchardef\beta="710C
\mathchardef\pi="7119
\mathchardef\sigma="711B
\mathchardef\lambda="7115
\mathchardef\mu="7116
\mathchardef\nabla="7272
\parskip=8pt plus3pt
\interlinepenalty=1500
\def\boxit#1{\vbox{\hrule\hbox{\vrule\kern1pc
   \vbox{\kern1pc#1\kern1pc}\kern1pc\vrule}\hrule}}
\def\sqr#1#2{{\vcenter{\vbox{\hrule height.#2pt
   \hbox{\vrule width.#2pt height#1pt  \kern#1pt
      \vrule width.#2pt}
     \hrule height.#2pt }}}}

 \leftline{\bf WAVEPACKET SOLUTIONS OF THE KLEIN-GORDON
 EQUATION  \hfil\break  }

\rightline{Shaun N Mosley,\footnote{${}^*$} {E-mail:
shaun.mosley@ntlworld.com  }
Sunnybank, Albert Road, Nottingham NG3 4JD, UK }

\beginsection Abstract

We present dispersion-free wavepacket solutions
$ \Psi_{ {\bf v} } $ to the
Klein-Gordon equation, with the only free parameter being the
wavepacket velocity $ {\bf v} \, .$ The $ \Psi_{ {\bf v} } $
are eigenvectors of a velocity operator $ {\bf \tilde v} \, $
with commuting components, which is symmetric in a certain
scalar product space. We show that $ {\bf \tilde v} \, $
corresponds to a classical generator
$ {\bf v} ( {\bf x} , {\bf k} ) \, $ which with its conjugate
 $ {\bf z} ( {\bf x} , {\bf k} ) \, $  may be obtained from
$ {\bf x} , {\bf k} \, $ by a canonical transformation.

\beginsection 1 Introduction

A particle with momentum $ ( m \gamma v ) \, $ in the $ z \, $
direction is usually represented by the plane wave
$$ \psi = A \, \exp  [  -  i \, m \gamma (t - v z ) ] \,
\, ,
  \qquad | v | < 1 , \, \gamma \equiv ( 1 - v^2 )^{- 1/2} \, $$
solution of the Klein-Gordon (KG) equation
$$  ( \partial_t^2 - {\bx \nabla}^2 + m^2 ) \, \psi
   = 0 \,   \eqno (1)  $$
(we use natural units with $\hbar = c = 1 \, ).$
 The plane wave is spread over all space, contrary to our usual
 experience. We explore
 a class of solutions to the KG equation that are localised,
 their magnitude approximately inversely
 proportional
 to the distance from the centre of the wavepacket; they travel
 without dispersion, and they are eigenfunctions of a particular
 velocity operator found below. The aim of this paper is to see
 if one can replace the usual plane wave basis states with those
 discussed here.
 
 The KG equation
admits the following solutions (considering the wavepacket velocity
$ v \, $ to be in the $ z \, $ direction for the moment)
$$ \eqalignno{
\Psi_{ v } ( {\bf x} , t ) &= \, { \sin ( m \gamma |v| {\tilde r} ) \over
m \gamma |v| {\tilde r} } \,
\exp [ - i \, m \gamma^2 (t - v z ) ] \, ,
  \qquad | v | < 1 , \, \gamma \equiv ( 1 - v^2 )^{- 1/2} \,  \cr
 &\equiv \, j_0 ( m \gamma |v| {\tilde r} )  \,
 \exp [  -  i \, m \gamma^2 (t - v z ) ]  ,
   & (2)  \cr
 \noalign{\noindent where \qquad \qquad \qquad
 $\tilde{r} \equiv \sqrt{ x^2 + \, y^2
  + \, \gamma^2 (z - v \, t )^2 } \, ,$}
   }  $$
 and $ j_0 \, $ is the spherical Bessel function of order zero.
 The wavepacket
 (2) can be thought of as a plane wave of wavelength
$ ( 2 \pi / m \gamma^2 |v| ) \, ,$
bounded by the
envelope function $ j_0 ( m \gamma |v| \tilde{r} )  \, $
  which has an extent of order one
  wavelength in the $ z \, $ direction (for
  non-relativistic velocities the envelope
  factor $ j_0 ( m \gamma |v| \tilde{r} ) \, $
  is nearly spherically symmetric about the point
  $ ( 0 , 0 , v t ) \, ).$ The wavepacket moves
  with velocity $ v \, $ in the $ z \, $ direction
  without dispersion. The
$ \Psi_{ v } \, $ solutions are a special case of the
wavepackets derived by MacKinnon. [1, 2]
That (2) satisfies the KG equation follows trivially
by applying the Lorentz coordinate transformation
 $$  z \rightarrow \gamma ( z - v t ) \, , \;
 x \rightarrow x \, , \; y \rightarrow y \, , \;
   t \rightarrow \gamma ( t - v z ) \, \eqno (3)  $$
 to the stationary KG solution
$$ \psi_{ v } = \, j_0 ( m \gamma |v| r ) \,
  \exp [  -  i \, m \gamma t  ]
  \, . \eqno (4) $$
The Lorentz transformation
is such that it exactly cancels out the negative $ z $
momentum components in the superposition making up (4),
as will be seen from the Fourier transform of
$ \Psi_{ v} \, $ discussed in the next section.
 
 Further solutions
$ \Psi_{ l ,v } \, $
of the KG equation with spin $ l \, $ can be generated from
$ \Psi_{ v } \, :$
$$ \Psi_{ l , v }  = ( \partial_x + \, i \, \partial_y )^l \,
 \Psi_{  v } \, = \, j_l ( m \gamma |v| \tilde{r} ) \,
  Y_{ l l } ( \theta , \phi ) \,
  \exp [ - i \, m \gamma^2 (t - v z ) ] \eqno (5) $$
   where
$ Y_{ l l } ( \theta , \phi ) \propto ( x + i y / r )^l \, $
   is the spherical harmonic.
  For wavepackets travelling in an arbitrary
direction with velocity $ {\bf v} \, $ then the
$ \Psi_v $ of (2) is
 $$ \eqalignno{
 \Psi_{{\bf v}} &= \, j_0 ( m \gamma |{\bf v}| {\tilde r} )  \,
 \exp [  -  i \, m \gamma^2 (t - {\bf v} \cdot {\bf x} ) ]
   & (6)  \cr
   \noalign{\noindent where now }
 \tilde{r} &\equiv \sqrt{ x_\perp^2
  + \, \gamma^2 ({\bf x}_\parallel - {\bf v} \, t )^2 } \, ,
   \qquad  {\bf x}_\perp \equiv {\bf x} \, - \,
  ({\bf \hat v} \cdot {\bf x}) \, {\bf \hat v} \, , \qquad
   {\bf x}_\parallel \equiv
  ({\bf \hat v} \cdot {\bf x}) \, {\bf \hat v} \, . \cr
     } $$
The (time dependant) distance $ \tilde{r} \, $
is zero at the wavepacket centre $ ( t , {\bf v} t ) \, $.
The $\Psi_{{\bf v}} \,  $ satisfy the identity
$$ i \, ( \partial_t + {\bf v} \cdot {\bx \nabla} \, ) \,
\Psi_{{\bf v}} \, = \, m \, \Psi_{{\bf v}} \eqno (7) $$
noting that
the operator $ ( \partial_t + {\bf v} \cdot {\bx \nabla}  )
\, $ commutes with $ \tilde{r} \; $ in (6).

We are curious as to whether $ \Psi_{\bf v} , \, \Psi_{\bf v'} $
are orthogonal, the natural approach to this question is to
enquire if  $ \Psi_{\bf v} \, $ is an eigenfunction of some
operator $ \tilde{\bf v} $ which has commuting components. It
turns out that this operator is easier to spot in Fourier
transform space.

 \beginsection 2 The Fourier transform of $ \Psi_{\bf v} \, $

 Positive energy solutions of the KG equation are of the form
 (see for example Schweber [3])
 $$ \eqalignno{
 \psi ( {\bf x} , t )
  &= {2 \over ( 2 \pi )^{3/2} } \int \phi ( k^0 , {\bf k} ) \,
  e^{ - i k^0 t + i {\bf k} \cdot  {\bf x} }
  \delta ( {k^0}^2 - {\bf k}^2 - m^2 ) \, \theta ( k^0 )  \,
  d^4 k \, ,  \cr
  &= {1 \over ( 2 \pi )^{3/2} } \int \phi ( {\bf k} ) \,
  e^{ - i \sqrt{m^2 + k^2 } t + i {\bf k} \cdot  {\bf x} }   \,
  {d {\bf k} \over  \sqrt{m^2 + k^2 } } \, , \qquad \qquad
  k \equiv | {\bf k}| \, ,  \cr
  &\equiv {\cal F}_{\bf k} \, \Big[ \phi ( {\bf k} ) \,
  { e^{ - i \sqrt{m^2 + k^2 } t  } \over \sqrt{m^2 + k^2 } }
  \, \Big] \, ({\bf x}) \, .  \cr
   }   $$
   We will call this $ \phi ( {\bf k} ) \, $ the momentum space
   counterpart to the $ \psi ( {\bf x} , t ) \, $
   wavefunction.
 In the first line above the step function
  $ \theta ( k^0 )  \, $ ensures that the energy
  $ k^0 = + \sqrt{m^2 + k^2} \, $ is positive, and
  $ {\cal F} \, $ is the usual 3-D Fourier transform. As
  $ \phi (  {\bf k} ) \, $ is a Lorentz scalar
  (due to the delta function), if we can find
  the $ \phi (  {\bf k} ) \, $ corresponding to the
  stationary solution (4) it will be straightforward to find the
  $ \phi (  {\bf k} ) \, $ corresponding to (2).
  Starting with the formula for the 3-D Fourier transform of
  spherically symmetric functions $ \phi ( k ) \, ,$ which is
  $ {\cal F}_{\bf k} \, \big[ \phi ( k ) \big] \,
  ({\bf x}) = \sqrt{ 2 \over \pi } \, { 1 \over r } \,
  \int \phi ( k ) \, \sin ( k r ) \, k \, dk  \, ,$ then
  $$ \eqalignno{
   &{\cal F}_{\bf k} \, \Big[ \sqrt{ \pi \over 2 } \,
   { \delta ( k - m \gamma  |v| ) \over ( m \gamma  v )^2 }
   \Big] \, ({\bf x})
 =  \, { \sin ( m \gamma |v| r ) \over
  m \gamma |v| r }  \equiv j_0 ( m \gamma |v| r ) \cr
  \noalign{ \noindent or using the delta function identity
  $ \delta ( k - m \gamma  |v| ) = \, |v| \,
   \delta ( \sqrt{m^2 + k^2 } - m \gamma  ) \, $   }
     &{\cal F}_{\bf k} \, \Big[ \Big\{ \sqrt{ \pi \over 2 } \,
  { \delta ( \sqrt{m^2 + k^2 } - m \gamma  )
 \over  m \gamma |v| } \Big\} \,  { e^{ - i \sqrt{m^2 + k^2 } t  }
 \over \sqrt{m^2 + k^2 } }\Big]
  =  \,  j_0 ( m \gamma |v| r ) \,
  \exp [  -  i \, m \gamma t  ]  \cr
     } $$
     which RHS is (4), so that the quantity in braces is the
    momentum space counterpart $ \phi ({\bf k}) \, $ to (4).
 Next we substitute into the
 $$ \phi ({\bf k}) \, = \, \sqrt{ \pi \over 2 } \,
  { \delta ( \sqrt{m^2 + k^2 } - m \gamma  )
 \over  m \gamma |v| } \eqno (8) $$
  the transformations
   $$\eqalign{
   &k_3 \rightarrow \gamma k_3 - \gamma v \sqrt{m^2 + k^2 } \, ,
   \qquad  k_1 \rightarrow  k_1 \, , \qquad
   k_2 \rightarrow k_2 \, , \; \cr
   &\sqrt{m^2 + k^2 } \rightarrow \gamma \sqrt{m^2 + k^2 }
  + \gamma v k_3  \; \cr
     } \eqno (9) $$
     corresponding to the Lorentz transformations (3),
  then we find
    $$\eqalignno{
   \phi (  {\bf k} ) \rightarrow \phi_v (  {\bf k} )
   &= \sqrt{ \pi \over 2 } \,
  { \delta ( \gamma \sqrt{m^2 + k^2 } - m \gamma - \gamma v k_3  )
 \over  m \gamma |v| } \cr
 &= \sqrt{ \pi \over 2 } \,
  { \delta ( \sqrt{m^2 + k^2 } - m -  v k_3  )
 \over  m \gamma^2 |v| }   \cr
 \noalign{ \noindent and so the $ \Psi_{ v } ( {\bf x} , t ) \, $
 of (6) is   }
 \Psi_{ v } ( {\bf x} , t ) &= {\cal F}_{\bf k} \,
 \Big[ \phi_v ( {\bf k} ) \,
  { e^{ - i \sqrt{m^2 + k^2 } t  } \over \sqrt{m^2 + k^2 } }
  \, \Big] \, ({\bf x}) \, \cr
  &= {\cal F}_{\bf k} \,
  \Big[ \sqrt{ \pi \over 2 } \, {1 \over  m \gamma^2 |v| } \,
  \delta ( \sqrt{m^2 + k^2 } - m -  v k_3  ) \,
  { e^{ - i \sqrt{m^2 + k^2 } t  } \over \sqrt{m^2 + k^2 } }
  \, \Big] \, ({\bf x}) \,  & (10)
   }  $$
   and the quantity in square brackets is the desired Fourier
   transform of $ \Psi_{ v } \, $ which we will call
   $ \alpha_{ v } ({\bf k}) \, :$
   $$ \alpha_{ v } ({\bf k})= \, \sqrt{ \pi \over 2 } \,
   {1 \over  m \gamma^2 |v| } \,
  \delta ( \sqrt{m^2 + k^2 } - m -  v k_3  ) \,
  { e^{ - i \sqrt{m^2 + k^2 } t  } \over \sqrt{m^2 + k^2 } }
  \, $$
   The delta function $ \delta (  \sqrt{ m^2 + \, k^2 }
  - \,  m \,    - \,  v  k_3 ) \, $ within
  $ \alpha_{ v } ({\bf k}) \, $ is non-zero on the surface
 $$ k_1^2 + k_2^2 + \gamma^{-2} [ k_3
     - m v \gamma^2 ]^2 = m^2 v^2 \gamma^2 \, ,  $$
     which is an ellipsoid tangential to the $ k_3 = 0 \, $
 plane at the origin, with centre $ ( 0 , 0 , m v \gamma^2 ) \, $
   and elongated in the $ k_3 \, $ direction. As already
   mentioned, the $ \alpha_{ v } ({\bf k}) \, $ has no negative
   $ k_3 \, $ component (the ellipsoid surface has
  $ k_3 \ge 0 \, ).$ For the general case when the velocity
  $ {\bf v} \, $ is in an arbitrary direction then
  $ \alpha_{\bf v } ({\bf k}) \, $
  is
  $$  \alpha_{\bf v } ({\bf k})
  =  \, \sqrt{ \pi \over 2 } \,
  {1 \over  m \gamma^2 |v| } \,
   \delta ( \sqrt{m^2 + k^2 } - m - {\bf v} \cdot {\bf k} ) \,
  { e^{ - i \sqrt{m^2 + k^2 } t  } \over \sqrt{m^2 + k^2 } }
   \, . \eqno (11)  $$
   Note that due to the delta function
   $$ \sqrt{m^2 + k^2 } = m \, + \, {\bf v} \cdot {\bf k}  $$
   which corresponds to (7), and the time evolution operator is
 $  e^{ - i \sqrt{m^2 + k^2 } t  }
 =  e^{ - i ( m + {\bf v} \cdot {\bf k} ) t  } $ where the
 $  e^{ - i  {\bf v} \cdot {\bf k}  t  } $ factor generates the
 translation $ {\bf v}  t \, $ as expected.

  \beginsection 3 The velocity operator

  Our goal is to find the velocity operator
  $ \tilde{\bf v} $ in Fourier transform
  space such that $ \tilde{\bf v} \, \alpha_{\bf v} ({\bf k})
  = \, {\bf v} \, \alpha ( {\bf k} ) \, .$
  Once we have found the operator at time $ t = 0 \, $ then the
  operator $ \tilde{\bf v}_t $ for time $ t \, $ will be
  $$ \tilde{\bf v}_t = e^{ - i \sqrt{m^2 + k^2 } t }
 \tilde{\bf v}_0  e^{ i \sqrt{m^2 + k^2 } t } \,
 . $$
  At time $ t = 0 \, $ then
  $ \alpha_{\bf v} ({\bf k}) $ is  (disregarding a factor)
 $$ \eqalignno{
 \alpha ( {\bf k}  )
 &= \, { \delta ( \sqrt{m^2 + k^2 } - m
  - {\bf v} \cdot {\bf k}  ) \over \sqrt{m^2 + k^2 } } \, .
  & (12) \cr
    }   $$
  Clearly the operator $ \tilde{\bf v} $ cannot be simply
  multiplication
  by a function of $ {\bf k} $ as the delta function is a 2-D
  surface, not a point.
  Differentiating $ \alpha_{\bf v} ({\bf k}) $ with respect to
  $ {\bf k} \, $ yields a term $ {\bf v} \, $ multiplied by the
 derivative of the delta function. We can avoid this complication
 by considering first the `anti-derivative' of the delta function
    which is a unit step function: we define
   $ E \, ( \chi ) = - 1/2 , \chi < 0 ; \;
    E \, ( \chi ) = 1/2 , \chi > 0 \, .$
  Then
  $$ E (  \sqrt{ m^2 + \, k^2 }
   - \, m \, - \,  {\bf v} \cdot {\bf k} )  \, $$ has the
   value $ - 1/2 \, $ within the delta function ellipsoid,
   and is $ 1/2 \, $ outside. Then
 $$ \eqalignno{
  {\bx \nabla}_k \, E (  \sqrt{ m^2 + \, k^2 }
     - \, m \, - \,  {\bf v} \cdot {\bf k} )  \,
 &=  \big( { {\bf k} \over \sqrt{ m^2 + \, k^2 } } \, - \, {\bf v}
   \big) \, \delta (  \sqrt{ m^2 + \, k^2 }
     - \, m \, - \,  {\bf v} \cdot {\bf k} )  \,   \cr
     &=  \big(  {\bf k}  \, - \, \sqrt{ m^2 + \, k^2 } \, {\bf v} \,
   \big) \, \alpha ( {\bf k} )  \, & (13)   \cr
 \noalign{\noindent and }
    ( {\bf k} \cdot {\bx \nabla}_k ) \,
 E (  \sqrt{ m^2 + \, k^2 }
     - \, m \, - \,  {\bf v} \cdot {\bf k} )  \,
 &=  \Big(  k^2 \,
   - \, ({\bf v} \cdot {\bf k}) \, \sqrt{ m^2 + \, k^2 } \Big) \,
 \alpha ( {\bf k} )   \cr
 &= \, m \,
   \Big( \sqrt{ m^2 + \, k^2 } - m  \,
    \Big) \, \alpha ( {\bf k} ) \cr
 \noalign{\noindent so that  }
 E (  \sqrt{ m^2 + \, k^2 }
     - \, m \, - \,  {\bf v} \cdot {\bf k} )  \,
 &= \, m \, {( {\bf k} \cdot {\bx \nabla}_k )}^{- 1} \,
 ( \sqrt{ m^2 + \, k^2 } - m ) \;
    \alpha ( {\bf k} )  \, . & (14)  \cr
     }  $$
  Now we substitute this last result into (13)
   obtaining
  $$ \eqalignno{
  \Big\{  {\bf k}  \,
 &- \, m \, {\bx \nabla}_k
 {( {\bf k} \cdot {\bx \nabla}_k )}^{- 1} \,
 \big(  \sqrt{ m^2 + \, k^2 } - m  \,
    \big) \, \Big\} \,  \alpha ( {\bf k} )   \,
  = \sqrt{ m^2 + \, k^2 } \, {\bf v} \; \alpha ( {\bf k} ) \, \cr
  \noalign{\noindent or  }
 \tilde{\bf v}_k  \,  \alpha ( {\bf k} )  \,
 \equiv \, \Big\{ { {\bf k} \over \sqrt{ m^2 + \, k^2 } } \,
 &- \, m \, { 1 \over \sqrt{ m^2 + \, k^2 } } \, {\bx \nabla}_k
 {( {\bf k} \cdot {\bx \nabla}_k )}^{- 1} \,
 \big(  \sqrt{ m^2 + \, k^2 } - m  \,
    \big) \, \Big\} \;  \alpha ( {\bf k} )   \,
        = {\bf v}  \;  \alpha ( {\bf k} )   \, .
        & (15) \cr    }   $$

 The operator $  {\bx \nabla}_k
 {( {\bf k} \cdot {\bx \nabla}_k )}^{- 1} $
 within $ \tilde{\bf v}_k  \, $ is defined as follows:
 $$  {\bx \nabla}_k
 {( {\bf k} \cdot {\bx \nabla}_k )}^{- 1}
 = \, {\bx \nabla}_k
 {( k \partial_k )}^{- 1}
 = \,(  - \, i \,  {\bx \nabla}_k k )
  ( - \, i \, \partial_k k)^{- 1} {1 \over k }
 \equiv \,(  - \, i \,  {\bx \nabla}_k k )
  D_k^{-1} \, {1 \over k }  \eqno (16) $$
 where $ D_k \, $ is the dilation operator
  $$ D_k \equiv \, - \, i \, ( \partial_k k ) \, $$
 which commutes with any operator of homogeneity degree zero
     (i.e. any operator invariant under a dilation of
     $ {\bf k} \, ).$ And
 $$ D_k^{-1} \, \phi ( {\bf k} )
 \equiv \, i \, ( \partial_k k)^{- 1} \phi ( {\bf k} )
 \equiv \, {i \over 2}  \, \Big[
 \int_0^1 \phi ( \lambda {\bf k} ) \, d \lambda \,
 - \, \int_1^\infty \phi ( \lambda {\bf k} ) \, d \lambda \,
    \Big] \,  \eqno (17) $$
    which is a right and left inverse of $ D_k \, .$
 The operator from (16) which is
 $ (  - \, i \,  {\bx \nabla}_k k )  D_k^{-1} \, {1 \over k }
 = D_k^{-1} \, ( - i {\bx \nabla}_k ) \, $
 can be shown to be symmetric
 in the usual 3-D inner product space, as
 $$ \eqalignno{
 &( - i {\bx \nabla}_k )^{\dag} = ( - i {\bx \nabla}_k ) \, ,
 \qquad D_k^{\dag} = k D_k {1 \over k} \, ,
  \qquad ( D_k^{- 1} )^{\dag} = k D_k^{- 1} {1 \over k} \, , \cr
 &[ D_k^{- 1} ( - i {\bx \nabla}_k ) ]^{\dag}
 = ( - i {\bx \nabla}_k )^{\dag} \, ( D_k^{- 1} )^{\dag}
 = ( - i {\bx \nabla}_k ) k D_k^{- 1} {1 \over k} \,
 = \, D_k^{- 1} ( - i {\bx \nabla}_k ) \, .  & (18) \cr
     }  $$

 From (15,18) it follows that $ \tilde{\bf v}_k $ is symmetric
 in the scalar product space
 $$ \Big\langle  \alpha_1 ({\bf k}) \, , \,  \alpha_2 ({\bf k})
 \Big\rangle_{w} \equiv
 \Big\langle \alpha_1 ({\bf k}) \, \big| \,
 ( \sqrt{ m^2 + \, k^2 } - m ) \, \sqrt{ m^2 + \, k^2 } \,
 \big| \, \alpha_2 ({\bf k}) \Big\rangle \,
 \eqno (19) $$
 instead of the usual Klein-Gordon scalar product space
 $ \Big\langle \alpha_1 ({\bf k}) \, \big|  \,
 \sqrt{ m^2 + \, k^2 } \,
 \big| \, \alpha_2 ({\bf k}) \Big\rangle \, .$
  As the components of $ \tilde{\bf v}_k $ which are
 $$  \{ \tilde{v}_{k1} \, , \,
 \tilde{v}_{k2} \, , \, \tilde{v}_{k3} \, \} \, \eqno (20) $$
 have a common
 eigenvector $ \alpha ({\bf k}) \, ,$ it also follows that these
 components commute:
we explicitly calculate the commutator
 $ [  \tilde{v}_{k1} \, , \, \tilde{v}_{k2} ] \, $ in the
 Appendix and check that it is zero.
 It follows that eigenvectors with different eigenvalues of
  $ \tilde{\bf v}_k $ are orthogonal to each other with respect
  to the scalar product space (19), i.e if
  we label $ \alpha_{{\bf v}_1} ({\bf k})  $ with the property
  $  \tilde{\bf v}_k \alpha_{{\bf v}_1}
  = {\bf v}_1 \alpha_{{\bf v}_1} $ etc, then
  $$ \langle \alpha_{{\bf v}_1} ({\bf k}) \, , \,
  \alpha_{{\bf v}_2} ({\bf k})
 \rangle_{w} = 0 \, , \qquad ( {\bf v}_1 \neq {\bf v}_2 ) \, . $$

 In configuration space the scalar product corresponding
 to (19) is
   $$ \Big\langle \psi_1 ({\bf x}) \, \big| \,
  - \partial_t^2 - \, i  \, m \, \partial_t \,
 \big| \, \psi_2 ({\bf x}) \Big\rangle \, \equiv  {1 \over 2} \;
 \int [ \partial_t \psi_1^* ({\bf x}) ] \;
 [ \, ( \partial_t + i \, m ) \psi_2 ({\bf x}) ] \; \, d^3 {\bf x}
 \qquad + \quad \hbox{s.c. }
 \eqno (21) $$
 where s.c. stands for the symmetric conjugate term. The operator
 corresponding to
$ \tilde{\bf v}_k $ is
$ \tilde{\bf v} = {\cal F}  \tilde{\bf v}_k {\cal F}^{-1}  $ ,
 found by substituting
$$ \eqalign{
 {\bf k} &\rightarrow - i {\bx \nabla} \, , \quad
 {\bx \nabla}_k \rightarrow - i {\bf x} \, , \quad \cr
 \sqrt{ m^2 + \, k^2 }  &\rightarrow  \sqrt{ m^2 - \, \nabla^2 }
 = \, i \, \partial_t  \, , \quad \cr
 {( {\bf k} \cdot {\bx \nabla}_k )} &\rightarrow
 - \, {( {\bx \nabla} \cdot {\bf x} )} = - \, {1 \over r^2 }
 ( \partial_r r ) r^2 \, \equiv - \, i \, {1 \over r^2 }
 D \, r^2 \, \cr
     } $$
 where $ D \, $ is the dilation operator $ D \equiv ( - i \,
 \partial_r r ) \, $ into (15), obtaining
 $$ \eqalign{
  \tilde{\bf v}
 &= \, - \, i \, { {\bx \nabla} \over \sqrt{ m^2 - \nabla^2 } } \,
 - \, i \,  m \, { 1 \over \sqrt{ m^2 - \nabla^2 } } \, {\bf x}
 \big[ {1 \over r^2 } ( i \, D )^{- 1} r^2  \big] \,
 ( \sqrt{ m^2 - \nabla^2 } - m  \, ) \,  \cr
 &= \, ( i \partial_t )^{-1}
 \Big( - \, i \,  {\bx \nabla}  \,
 - \,  m \, {\bf x}
 \big[ {1 \over r^2 } ( D )^{- 1} r^2  \big] \,
 (  i \partial_t - m  \, ) \Big) \, \cr
   }  \eqno (22) $$
   where $  ( D )^{- 1} $ is defined as the space counterpart to
 (17). Recalling the $ \Psi_{\bf v} $ of (6):
 $$ \Psi_{{\bf v}} =  \, j_0 ( m \gamma |{\bf v}| {\tilde r} ) \,
 \exp [  -  i \, m \gamma^2 (t - {\bf v} \cdot {\bf x} ) ]
  $$
 then the result (22) implies that at time $ t = 0 \, $
   $$ \eqalign{
  ( {\bx \nabla} + {\bf v} \, \partial_t ) \,
  \Psi_{\bf v} ( {\bf x} )
 &= \, i \,  m \, {\bf x}
 \big[ {1 \over r^2 } ( D )^{- 1} r^2  \big] \,
 (  i \partial_t - m  \, ) \, \Psi_{\bf v} ( {\bf x} ) \, , \cr
   }   $$
   which may be verified by explicit calculation, each side
   having the value
   $$ - m \gamma |v| \, {{\bf x} \over {\tilde r} } \,
    j_1 ( m \gamma |v| {\tilde r} )  \,
 \exp [ - i \, m \gamma^2 (t - {\bf v} \cdot {\bf x} ) ] \, . $$
 where $ j_1 \, $ is the spherical Bessel function of order one.

\beginsection 4 A canonical transformation from variables
 $ ( {\bf x} , {\bf k} )  \rightarrow \,
 ( {\bf z} , {\bf v} )  \, $
 
 The fact that the components of the velocity operator
 commute leads to inquire whether there is a canonical transformation
 underlying this property. Classically the $ {\bf \tilde{v}} \, $
 operator of (22) corresponds to the generator
 $$ {\bf v} ( {\bf x} \, , \, {\bf k} )
  =  \, { 1 \over \sqrt{ m^2 + \, k^2 } } \,
  {\bf k} \,
 - \, m \, { \big(  \sqrt{ m^2 + \, k^2 } - m  \, \big) \over
 \sqrt{ m^2 + \, k^2 } ( {\bf k} \cdot {\bf x} ) } \,
\, {\bf x}  \,\, $$
and it may be verified that the Poisson bracket
$$     \{ v_a \, , \,  v_b \} = 0 \qquad \qquad \qquad
a,b = 1,2,3 $$
where the Poisson bracket (P.b.) between
two variables $ f( {\bf x } , {\bf k } )  \, , \,
g( {\bf x } , {\bf k }  ) $ is defined as
$$  \{ f,g \} \equiv
 {\left( \partial f / {\partial {\bf x}}\right) \cdot
 \left( \partial g / {\partial {\bf k}}\right)}
 - {\left( \partial f / {\partial {\bf k}}\right) \cdot
 \left( \partial g / {\partial {\bf x}}\right)} \, .  $$
 The generator conjugate to $ {\bf v} \, $ is
 $$ {\bf z} ( {\bf x} \, , \, {\bf k} )
 = \, { \sqrt{ m^2 + \, k^2 }
 ( {\bf k} \cdot {\bf x} )
  \over m \, \big( \sqrt{ m^2 + \, k^2 } - m  \, \big) } \,
  {\bf k} $$
  which means that
  $$  \{ z_a \, , \,  v_b \} = \, \delta_{ab} \,
  , \qquad \{ z_a \, , \,  z_b \} = \, 0 $$
  as may be verified by labourious calculation.
  So then the transformation from the variables
  $ ( {\bf x} , {\bf k} )  \, $ to new variables
  $ ( {\bf z} , {\bf v} )  \, $ is canonical:
      $$ \cases{
  {\bf z} ( {\bf x} \, , \, {\bf k} )
 = \displaystyle \, { \sqrt{ m^2 + \, k^2 }
 ( {\bf k} \cdot {\bf x} )
  \over m \, \big( \sqrt{ m^2 + \, k^2 } - m  \, \big) } \,
  {\bf k} &  \cr
  {\bf v} ( {\bf x} \, , \, {\bf k} )
  = \displaystyle \, { 1 \over \sqrt{ m^2 + \, k^2 } } \,
  {\bf k} \,
 - \, m \, { \big(  \sqrt{ m^2 + \, k^2 } - m  \, \big) \over
 \sqrt{ m^2 + \, k^2 } ( {\bf k} \cdot {\bf x} ) } \,
\, {\bf x}  \, & \cr
     }  \eqno (23) $$
    with inverse
    $$ \cases{
  {\bf x} ( {\bf z} \, , \, {\bf v} )
 = \displaystyle \,
 { 1 \, - \, (  {\bf \hat z} \cdot {\bf v} )^2  \over m }  \,
 \Big( {1 \over 1 \, + \, ( {\bf \hat z} \cdot {\bf v} )^2 } \,
 {\bf z} \, - \, { z^2 \over 2 \, ( {\bf z} \cdot {\bf v} ) } \,
  {\bf v} \Big) \,  & \cr
  {\bf k}  ( {\bf z} \, , \, {\bf v} )
  = \displaystyle \, { 2 \, m  \, ( {\bf \hat z} \cdot {\bf v} )
  \over 1 \, - \, ( {\bf \hat z} \cdot {\bf v} )^2 } \,
 {\bf \hat z} \, .& \cr
     }  \eqno (24) $$
     where $ {\bf \hat z} \equiv {\bf z} / z \, .$
     The value of $ | v | \, $ from (23) is not limited to be
     less than unity, however the canonical transformation
     is meaningless if $ | v | \, $ is allowed to equal unity,
     because the denominator
     $\big( 1 \, - \, ( {\bf \hat z} \cdot {\bf v} )^2 \big) $
     in (24) can then be zero.
  The Hamiltonian in the new coordinates is
  $$ \sqrt{ m^2 + \, k^2 }
   = \, m \, \Big( {  1 \, + \, (  {\bf \hat z} \cdot {\bf v} )^2
  \over  1 \, - \, (  {\bf \hat z} \cdot {\bf v} )^2 }
  \Big) \, . $$

    \beginsection 5 Orthogonality relation

     We calculate orthogonality relations between coinciding
     wavepackets with differing velocities in the $  z \, $
     direction
     $ \Psi_v , \, \Psi_{v'} \, ,$ or equivalently
  between their momentum space counterparts
  $ \alpha_v , \, \alpha_{v'} \, ,$ recalling from (11)
  that at time $ t = 0 \, $
  $$ \alpha_v   =  \, \sqrt{ \pi \over 2 } \,
  {1 \over  m \gamma^2 |v| \sqrt{m^2 + k^2 } } \,
    \delta ( \sqrt{m^2 + k^2 } - m - v \, k_3 )  \, .  $$
   We have mentioned that
  when $ v \neq v' \, $ the surface delta functions inside
   $ \alpha_v , \, \alpha_{v'} \, $  only touch at the
  origin, and so whether they are orthogonal or not will depend
  on whether there are any factors of $ k \, $ multiplying the
  delta functions.
   We first write the above in spherical coordinates
 $$ \eqalignno{
  \alpha_v
  &=  \, \sqrt{ \pi \over 2 } \, {1 \over  m \gamma^2 |v|
  \sqrt{m^2 + k^2 } } \,
    \delta ( \sqrt{m^2 + k^2 } - m - v \, k \, \cos \theta )  \cr
  &=  \, \sqrt{ \pi \over 2 } \, {1 \over  m \gamma^2 |v|
  v k \sqrt{m^2 + k^2 } } \,
  \delta \big(  \cos \theta  \, - \, { \sqrt{ m^2 + \, k^2 }
     - \, m \over v k }   \big)  \, .\cr
    }   $$
Then recalling the scalar product space
$ \langle  \cdot \, , \,  \cdot \rangle_{w} \, $ of (19)
$$ \eqalignno{
 \Big\langle  \alpha_v  \, , \,  \alpha_{v'}  \Big\rangle_{w}
&= \Big\langle \alpha_v \, \big| \,
 ( \sqrt{ m^2 + \, k^2 } - m ) \, \sqrt{ m^2 + \, k^2 } \,
 \big| \, \alpha_{v'}  \Big\rangle \, \cr
&\equiv \int ( \sqrt{ m^2 + k^2 } - m ) \, \sqrt{ m^2 + \, k^2 } \,
\alpha_v^*  \alpha_{v'} \;
  d( \cos \theta_k ) \, k^2 dk \, d\phi \cr
&= \, {\pi^2 \over m^2 \gamma^2 \gamma'^2 |v v'| v v' } \,
\int \,  {( \sqrt{ m^2 + k^2 } - m ) \,
\sqrt{ m^2 + \, k^2 } \over   k^2 (m^2 + k^2) } \;
   \delta \big(  \cos \theta  \, - \,
    { \sqrt{ m^2 + \, k^2 } - \, m \over v k } \big) \; \cr
 &\qquad \qquad \times  \delta \big(  \cos \theta  \, - \,
    { \sqrt{ m^2 + \, k^2 } - \, m \over v' k } \big) \;
  d( \cos \theta ) \, k^2 dk \,  \cr
&= \, {\pi^2 \over m^2 \gamma^2 \gamma'^2 |v v'| v v' } \,
  \int_0^{k_{max}} \,  {( \sqrt{ m^2 + k^2 } - m ) \,
 \over   k^2 \sqrt{ m^2 + \, k^2 } } \;
   \delta \big(  { \sqrt{ m^2 + \, k^2 } - \, m \over v' k } \, - \,
    { \sqrt{ m^2 + \, k^2 } - \, m \over v k } \big) \;
     k^2 dk \,  \cr
 &= \, {\pi^2 \over m^2 \gamma^2 \gamma'^2 |v v'| v v' } \,
   \int_0^{k_{max}} \,  {( \sqrt{ m^2 + k^2 } - m ) \,
 \over  \sqrt{ m^2 + \, k^2 } }  \;
   \delta \left( \big({ v - v' \over v v' }\big)
    \big({ \sqrt{ m^2 + \, k^2 } - \, m \over  k }\big) \, \right) \;
   dk \,  \cr
    }   $$
    Note that
  $$ \delta  \big(
    { \sqrt{ m^2 + \, k^2 } - \, m \over  k } \big)
    =  2 \, m \, \delta ( k ) \, , $$
    but this term is annihilated by the
    $ ( \sqrt{ m^2 + k^2 } - m ) \, $ factor in the integrand
    which is equal to $ k^2 / 2 m \, $ as $ k \rightarrow 0 \, .$
 Hence we can regard $ ( v - v' ) \, $ as the subject of the
 delta function and pull out the remaining factor obtaining
 $$ \eqalignno{
 \Big\langle  \alpha_v  \, , \,  \alpha_{v'}  \Big\rangle_{w}
 &= \, {\pi^2 \over m^2 \gamma^2 \gamma'^2 |v v'| v v' } \,
  \int_0^{k_{max}} \,  {k \over  \sqrt{ m^2 + \, k^2 } } \;
   \delta  \big({ v - v' \over v v' }\big) \,  \;   dk \,  \cr
   &= \, {\pi^2 \over m^2 \gamma^4 v^2  } \,
   \delta  \big( v - v' \big) \,
  \Big| \,  \sqrt{ m^2 + \, k^2 } \;   \Big|_0^{k_{max}}  \cr
 \noalign{\noindent and recalling that
  $ k_{max} = 2 m v \gamma^2  \, $ }
   \Big\langle  \alpha_v  \, , \,  \alpha_{v'}  \Big\rangle_{w}
   &= \, {\pi^2 \over m^2 \gamma^4 v^2  } \,
   \delta  \big( v - v' \big) \,
  \Big( { 1 + v^2 \over 1 - v^2 } m \, - \, m  \, \Big) \cr
   &= \, {\pi^2 \over m^2 \gamma^4 v^2  } \,
   \delta  \big( v - v' \big) \,
  \Big( 2 \, m \, v^2 \gamma^2  \Big) \cr
  &= \, { 2 \, \pi^2 \over m \gamma^2   } \,
   \delta  \big( v - v' \big) \, . \cr
    }   $$

 \beginsection 6 Coordinate transformation for
 $ \Psi_{ l, v } \rightarrow \,  \Psi_{ l, v' } \, $
 
We will show that there is a coordinate transformation which
 changes the $ \Psi_v \, $ of (2)
 $$\eqalignno{
 \Psi_v ( {\bf x} , t ) &=  \, j_0 ( m \gamma |v|
      \sqrt{ x^2 + \, y^2
  + \, \gamma^2 (z - v \, t )^2 } \,  ) \;
   \exp [ -  i \, m \gamma^2 (t - v z ) ] \, \cr
   \noalign{ \noindent to }
 \Psi_{v'} ( {\bf x} , t ) &=  \, j_0 ( m \gamma' |v'|
      \sqrt{ x^2 + \, y^2
  + \, \gamma'^2 (z - v' \, t )^2 } \,  ) \;
   \exp [ -  i \, m \gamma'^2 (t - v' z ) ] \, . \cr
    }  $$
 From inspection of the above we see that Lorentz transformations
 do not preserve the form of the  $ \Psi_v \, ,$ (as an
 example the spherical solution (3) is not of the form
  $ \Psi_v \, ).$
The following is the required transformation:
     $$ \cases{
  t \, =  t' & \cr
  z \, = \displaystyle{ { 1 \over \gamma^2 v } \,
  \Big( \gamma'^2 v' \, z'
    + \,[ \gamma^2 \, - \, \gamma'^2  ] \, t' \Big) \,
    =  { \gamma'^2 v' \over \gamma^2 v } \, z'\,
     + \, \Big( v \,
  - { \gamma'^2 v'^2 \over \gamma^2 v }  \Big) \, t' } &  \cr
  x \, = \displaystyle{ { \gamma' |v'| \over \gamma |v| } x' \, , \qquad
   y \, = { \gamma' |v'| \over \gamma |v| } y' \, } & \cr
     }  \eqno (25) $$
    It is readily checked that (25) implies
   $$ \eqalign{
  \gamma^2 ( t - v z )  \, &= \gamma'^2 ( t' - v' z' ) \cr
 \gamma^2 v ( z - v t ) \,
   &=  \gamma'^2 v' ( z' - v' t' ) \,   \cr
     }   $$
     as necessary.
 Remarkably under (25) the time coordinate remains unchanged,
 but the space coordinate transformations are specific to both
  the original and transformed velocities
 $ v $ and $ v' \, .$ So that if the transformation (25) is
 applied to, say, $ \Psi_{ l, v'' } \, $ with
 $ v'' \neq v \, ,$ the resulting function
 does not in general have the form $ \Psi_{ v } \, .$

 \beginsection 7 Outlook
 
The wavepacket solutions
 $ \Psi_{ {\bf v} } ({\bf x}, t )  \, $  of the KG equation
 are eigenvectors of the velocity operator
 (22), whose components commute and are symmetric in the
 inner product space (21). These wavefunctions are interesting
 alternative basis state to the usual plane waves.

 \beginsection  Appendix
 
Recalling
 $$ \tilde{\bf v}_k
 = \,  { {\bf k} \over \sqrt{ m^2 + \, k^2 } } \,
 - \, m \, { 1 \over \sqrt{ m^2 + \, k^2 } } \,
 D_k^{-1} \, ( - i {\bx \nabla}_k )
 \big(  \sqrt{ m^2 + \, k^2 } - m  \, \big) \,  $$
 with  $ D_k \equiv \, - \, i \, ( \partial_k k ) \, $ and
 $$ D_k^{-1} \, \phi ( {\bf k} )
 \equiv \, {i \over 2}  \, \Big[
 \int_0^1 \phi ( \lambda {\bf k} ) \, d \lambda \,
 - \, \int_1^\infty \phi ( \lambda {\bf k} ) \, d \lambda \,
 \Big] \, , $$
  we will need the following
 identities:
 $$\eqalignno{
  &\big[ D_k^{-1} \, , \, k \nabla_1 \big] \, = \, 0 \, & (A1) \cr
 &\nabla_1  D_k^{-1} = \, {1 \over k} ( k  \nabla_1 ) \, D_k^{-1}
 = \, {1 \over k}  D_k^{-1} k  \nabla_1  \, & (A2) \cr
 &\big[ D_k^{-1} \, , \, k_1 \nabla_2  - k_2 \nabla_1 \big] \,
 = \, 0 \, & (A3) \cr
 &\big[ D_k \, , \, { k \over \sqrt{ m^2 + \, k^2 } } \big] \,
 = \, - \, i \, { m^2 k \over {\sqrt{ m^2 + \, k^2 }}^3 } \,
   & (A4) \cr
 &\big[ D_k^{-1} \, , \, { k \over \sqrt{ m^2 + \, k^2 } } \big]
 \,
 = \, i \, D_k^{-1} { m^2 k \over {\sqrt{ m^2 + \, k^2 }}^3 } \,
 D_k^{-1} & (A5) \cr
    }  $$
    To calculate
    $ \big[  \tilde{v}_1 \, , \, \tilde{v}_2  \big] \, $
  we first calculate those parts of the commutator
  with the $ { {\bf k} \over \sqrt{ m^2 + \, k^2 } } \, $ terms,
  which are
     $$\eqalignno{
   \, - \, & \Big[   { k_1 \over \sqrt{ m^2 + \, k^2 } } \,
   \, , \, \, m \, { 1 \over \sqrt{ m^2 + \, k^2 } } \,
 D_k^{-1} \, ( - i \nabla_2 )
 \big(  \sqrt{ m^2 + \, k^2 } - m  \, \big) \,   \Big] \, \cr
  &\qquad + \,   \Big[   { k_2 \over \sqrt{ m^2 + \, k^2 } } \,
   \, , \, \, m \, { 1 \over \sqrt{ m^2 + \, k^2 } } \,
 D_k^{-1} \, ( - i \nabla_1 )
 \big(  \sqrt{ m^2 + \, k^2 } - m  \, \big) \,   \Big] \, \cr
 &= \, i \, { m \over m^2 + \, k^2  } \, ( k_1 \nabla_2
   - k_2 \nabla_1 ) \, k \, D_k^{-1} \,{ 1 \over k } \,
   \big(  \sqrt{ m^2 + \, k^2 } - m  \, \big) \,
    - \, i \, {m \over \sqrt{ m^2 + \, k^2 } } \,
    ( \nabla_1 k_2 -  \nabla_2 k_1 ) \, D_k^{-1} \,
 { \sqrt{ m^2 + \, k^2 } - m  \over \sqrt{ m^2 + \, k^2 } } \cr
 &= \, i \, m \, ( k_1 \nabla_2 - k_2 \nabla_1 ) \,
   \Big( \, { 1 \over \sqrt{ m^2 + \, k^2 } } \,
   \Big[  { k \over \sqrt{ m^2 + \, k^2 } }  \, , \, D_k^{-1} \,
     \Big] { 1 \over k } \,
   \big(  \sqrt{ m^2 + \, k^2 } - m  \, \big) \, \Big) \,
   \cr
   &= \, i \, m \, ( k_1 \nabla_2 - k_2 \nabla_1 ) \,
   \Big( \, { 1 \over \sqrt{ m^2 + \, k^2 } } \,
   \big( \, - \, i \, D_k^{-1} { m^2 k \over {\sqrt{ m^2 + \, k^2 }}^3 } \,
   D_k^{-1}  \big) { 1 \over k } \,
   \big(  \sqrt{ m^2 + \, k^2 } - m  \, \big) \, \Big) \, .& (A6)
   \cr
    }  $$
   Secondly we calculate how the second terms of
 $\tilde{\bf v}_k  \, $ commute with each other, yielding
  $$\eqalignno{
  &m^2 \,  { 1 \over \sqrt{ m^2 + \, k^2 } } \,
 D_k^{-1} \, \Big( ( - i \nabla_1 ) \,
 {  \sqrt{ m^2 + \, k^2 } - m   \over \sqrt{ m^2 + \, k^2 } } \,
  D_k^{-1} \,  ( - i \nabla_2 ) \,
   -  \, ( - i \nabla_2 ) \, {  \sqrt{ m^2 + \, k^2 } - m   \over \sqrt{ m^2 + \, k^2 } } \,
  D_k^{-1} \,  ( - i \nabla_1 ) \, \Big) \,
  ( \sqrt{ m^2 + \, k^2 } - m )  \,   \cr
 = &m^2 \,  { 1 \over \sqrt{ m^2 + \, k^2 } } \,
  D_k^{-1} \, \Big( \, - \,  i
 { m \, k_1 \over \sqrt{ m^2 + \, k^2 }^3 } \, ( - i \nabla_2 k ) \,
  D_k^{-1} { 1 \over k } \,
  + \, i
 { m \, k_2 \over \sqrt{ m^2 + \, k^2 }^3 } \, ( - i \nabla_1 k ) \,
  D_k^{-1} { 1 \over k } \, \Big) \,
  ( \sqrt{ m^2 + \, k^2 } - m )  \,
    \cr
    = &\, - \, m^2 \, ( k_1 \nabla_2 - k_2 \nabla_1 ) \,
     { 1 \over \sqrt{ m^2 + \, k^2 } } \,
 D_k^{-1} \, { m \, k \over \sqrt{ m^2 + \, k^2 }^3 } \,
  D_k^{-1} { 1 \over k } \, ( \sqrt{ m^2 + \, k^2 } - m )
    & (A7)  \cr
   }  $$
 Adding (A6) and (A7) yields zero.

\beginsection References

\item{[ 1 ]}   L. MacKinnon,
 Found. Phys. {\bf 8 }, 157 (1978)
\item{[ 2 ]}   L. MacKinnon,
Lett. Nuovo Cimento {\bf 31 }, 37 (1981)
\item{[ 3 ]}   S. S. Schweber, {\it An introduction to
relativistic quantum field theory} (Harper and Row, New York, 1961)
p57

\end